\documentstyle[12pt]{article}

\def\beq{\begin{equation}}
\def\eeq{\end{equation}}
\def\barr{\begin{eqnarray}}
\def\earr{\end{eqnarray}}

\def\b{\bigskip}

\def\cs{Chern-Simons }

\begin{document}

\title{Integer Quantization of the Chern-Simons Coefficient in a Broken
Phase\footnote{UCONN-94-8;hep-th/9411062}}

\author{\normalsize{Lusheng Chen, Gerald Dunne, Kurt Haller and Edwin
Lim-Lombridas} \\
\normalsize{Department of Physics}\\
\normalsize{University of Connecticut}\\
\normalsize{Storrs, CT 06269 USA}\\}

\date{October 28 , 1994}

\maketitle

\begin{abstract}
We consider a spontaneously broken nonabelian topologically massive gauge
theory in a broken phase possessing a residual nonabelian symmetry. Recently
there has been some question concerning the renormalization of the Chern-Simons
coefficient in such a broken phase. We show that, in this broken vacuum, the
renormalized ratio of the Chern-Simons coupling to the gauge coupling is
shifted by $1/4\pi$ times an integer, preserving the usual integer quantization
condition on the bare parameters.
\end{abstract}

\section{Introduction}
It is well known that the addition of a \cs term to the conventional Yang-Mills
gauge field action in $2+1$ dimensions leads to a massive transverse gauge
mode, with mass equal to the \cs coupling constant, $m$ \cite{jackiw1}. In a
nonabelian theory, consistency of the quantum theory under large gauge
transformations requires that the dimensionless ratio ${4\pi m\over g^2}$ be
quantized as an integer, where $g^2$ is the gauge coupling. This is most easily
understood as a field theoretical analogue of Dirac's monopole quantization
condition which requires ${\rm exp}\left(i({\rm action})\right)$ to be gauge
invariant \cite{dirac}. Pisarski and Rao have shown \cite{pisarski} that in a
1-loop perturbative analysis of topologically massive $SU(N)$
Chern-Simons-Yang-Mills gauge theory, the bare ratio $q\equiv{4\pi m\over g^2}$
receives a finite renormalization shift equal to the integer $N$. It has since
been shown that there is no further correction at two loops \cite{giavarini},
and there exist general arguments that this should be true to all orders
\cite{pisarski,giavarini}. On the other hand, one can also generate a gauge
field mass by the conventional Higgs mechanism, in a vacuum in which the gauge
symmetry is spontaneously broken. In this Letter we consider the perturbative
fate of the integer quantization condition on the \cs coupling in the broken
phase of a spontaneously broken nonabelian topologically massive theory. We
show that if there is still a nonabelian symmetry present in the broken phase,
then the one-loop renormalized ratio $\left[ {4\pi m\over g^2}\right]_{\rm
ren}$ is shifted from its bare value by an integer.

The abelian spontaneously broken \cs theory was investigated in
\cite{khlebnikov1,spiridonov}, where it was shown that in the broken vacuum
$q\equiv{4\pi m\over g^2}$ receives a finite renormalization shift which is a
complicated function of the various mass scales: the \cs mass $m$, the gauge
coupling $g^2$, and the symmetry breaking mass scale. A similar situation
exists for a completely broken nonabelian theory \cite{khlebnikov2}. However,
here there is no apparent contradiction with the \cs quantization condition
since there is no residual nonabelian symmetry in the broken vacuum. To probe
this question more deeply, the authors of Ref. \cite{khare1} introduced an
ingenious model in which a nonabelian topologically massive theory is partially
broken, leaving a nonabelian gauge symmetry in the broken phase. One's
immediate expectation is that once again the renormalized ratio $\left[ {4\pi
m\over g^2}\right]_{\rm ren}$ should be quantized, but with an integer
renormalization shift characteristic of the smaller unbroken residual gauge
symmetry. We have reconsidered this model both in a detailed canonical
treatment and, as reported in this Letter, in a direct perturbative analysis.
We disagree with the result reported in Ref. \cite{khare1}. We find that the
integer quantization condition {\bf is} indeed preserved to one loop. The
mechanism by which this quantization arises in a perturbative analysis is
considerably more involved than the Pisarski and Rao case \cite{pisarski} where
there is no symmetry breaking. The cancellations required to produce the
integer shift can be viewed as new `topological Ward identities' for the
spontaneously broken theory, generalizing the `topological Ward identity' found
in \cite{pisarski}.

For simplicity, we consider an $SU(3)$ topologically massive gauge theory,
coupled to a triplet of charged scalar fields with a potential possessing a
vacuum in which the $SU(3)$ symmetry is spontaneously broken to $SU(2)$. We
later show that our result applies also to the case of $SU(N)$ broken to
$SU(N-1)$, for $N\geq 3$.

The (Euclidean) Lagrange density for this model is
\barr
{\cal L}&={1\over 4}F^a_{\mu\nu }F^a_{\mu\nu }+i{m\over
2}\epsilon_{\mu\nu\rho }\left( \partial_\mu A^a_\nu A^a_\rho
+{1\over 3}gf^{abc}A^a_\mu A^b_\nu A^c_\rho\right)\cr
&+\left( D_\mu \Phi\right)^\dagger\left( D_\mu
\Phi\right)-\mu^2\left(\Phi^\dagger
\Phi\right)+\lambda\left(\Phi^\dagger\Phi\right)^2
\label{lag}
\earr
The gauge fields, $A_\mu=A_\mu^a T^a$, take values in the defining
representation. We use the hermitean generators $T^a={1\over 2}\lambda^a$,
satisfying $[T^a , T^b ]=i f^{abc} T^c$, where the $\lambda^a$ are the $SU(3)$
Gell-Mann matrices. We present our analysis in Euclidean space-time to permit
direct comparison with the results of Ref. \cite{pisarski}. The covariant
derivative is $D_\mu \Phi = \left( \partial_\mu - i g A_\mu \right ) \Phi$,
and the gauge curvature is
$F^a_{\mu\nu }=\partial_\mu A^a_{\nu}-\partial_\nu A^a_{\mu}+g
f^{abc}A^b_{\mu}A^c_{\nu}$.
Note that both the \cs coupling $m$ and the gauge coupling $g^2$ have
dimensions of mass. For definiteness we choose $m$ positive. The \cs term in
(\ref{lag}) is not invariant under large gauge transformations with nonzero
winding number, but the (Euclidean) action changes by an integer multiple of
$2\pi i$ provided the following quantization condition holds
\cite{jackiw1,pisarski}:
\barr
q\equiv {4\pi m\over g^2} = integer
\label{condition}
\earr
The Higgs potential in (\ref{lag}) has a nontrivial vacuum in which $\Phi$ has
the (tree approximation) vacuum expectation value
\barr
\Phi_0={1\over \sqrt{2}} \left(\matrix{0\cr 0\cr v}\right),\quad\quad
v\equiv\sqrt{\mu^2\over\lambda}
\label{vacuum}
\earr
In this nontrivial vacuum we express $\Phi$ as $\Phi=\Phi^\prime+\Phi_0$, where
\barr
\Phi^\prime={1\over \sqrt{2}} \left(\matrix{\psi_1+i\chi_1\cr\psi_2+i\chi_2\cr
\psi_3+i\chi_3}\right)
\label{field}
\earr
The Lagrange density (\ref{lag}) is supplemented with the `t Hooft gauge fixing
and ghost terms :
\barr
{\cal L}_g=-{1\over 2\xi}\left(\partial_\mu A^a_\mu -i g \xi \left(
\Phi_0^\dagger T^a \Phi-\Phi^\dagger T^a \Phi_0\right)\right)^2
+\partial_\mu\bar\eta^a\partial_\mu\eta^a -igf^{abc}\partial_\mu\bar\eta^a
A_\mu^b\eta^c
\label{ghost}
\earr
A simple calculation leads to the following quadratic gauge field portion of
the Lagrange density
\barr
{\cal L}_{\rm quad} ={1\over 2}\partial_\mu A_\nu^a (\partial_\mu
A_\nu^a-\partial_\nu A_\mu^a)+i{m\over 2}\epsilon_{\mu\nu\rho} \partial_\mu
A^a_\nu A^a_\rho -{1\over 2\xi}\left(\partial_\mu A^a_\mu\right)^2+{g^2\over 2}
\left(\Phi_0^\dagger \{T^a ,T^b \} \Phi_0\right) A_\mu^a A_\mu ^b
\label{quadratic}
\earr
and the anticommutation relations in (\ref{quadratic}) are conveniently
summarized as: $\{T^a , T^b \}={1\over 3}\delta^{ab}+ d^{abc}\,T^c$. We deduce
the following gauge field propagators (in the Landau gauge: $\xi=0$), diagonal
in color indices. In the unbroken sector (corresponding to gauge field
components $a=1,2,3$) the gauge field has the standard topologically massive
propagator:
\barr
\Delta_{\mu\nu}^{\rm unbroken}={(k^2 \delta_{\mu\nu}-k_\mu k_\nu)-m
\epsilon_{\mu\nu\rho} k_\rho\over k^2 (k^2+m^2)}
\label{unbroken}
\earr
The remaining gauge field components acquire mass through the \cs-Higgs
mechanism \cite{pisarski}, leading to a propagator with two mass poles:
\barr
\Delta_{\mu\nu}^{\rm broken}={(k^2 \delta_{\mu\nu}-k_\mu k_\nu)
(k^2+m_W^2)/k^2-m\epsilon_{\mu\nu\rho }k_\rho\over
(k^2+m_+^2)(k^2+m_-^2)}
\label{broken}
\earr
where the two mass poles are related to the \cs mass and the ``W-boson'' mass
$m_W$ by
\barr
m_\pm=\sqrt{m_W^2+{m^2\over4}}\pm{m\over 2}
\label{masses}
\earr
For the iso-doublet massive gauge fields (with $a=4,5,6,7$) the W-boson mass is
$m_W={vg/2}\equiv m_D$, while for the iso-singlet massive gauge field (with
$a=8$) the W-boson mass is $m_W={vg/\sqrt 3}\equiv m_S$. The most direct way
\cite{pisarski} to derive the broken propagators (\ref{broken}) is to invert
the self-energy in (\ref{quadratic}), but a more physical approach is to note
that the two masses $m_\pm$ correspond precisely to the two characteristic
frequencies $\omega_\pm$ of a planar quantum mechanical system in the presence
of a perpendicular magnetic field (corresponding to the \cs interaction) and a
harmonic potential (corresponding to the Higgs potential) \cite{dunne}.

The 3-gluon vertex differs from the standard QCD vertex by the inclusion of a
parity-odd \cs contribution, giving the combined vertex :
\barr
V_{\mu\nu\rho}^{abc}(p,q,r)=-i g f^{abc}\left[
(r-q)_\mu\delta_{\nu\rho}+(p-r)_\nu
\delta_{\mu\rho}+(q-p)_\rho\delta_{\mu\nu}-m\epsilon_{\mu\nu\rho}\right]
\label{vertex}
\earr
In addition there are the conventional vertices of spontaneously broken gauge
theory.

The gauge field self-energy can be expressed as
\barr
\Pi_{\mu\nu}= (p^2 \delta_{\mu\nu}-p_\mu p_\nu) \Pi_{\rm even} (p^2)
+m\epsilon_{\mu\nu\rho}p_\rho \Pi_{\rm odd} (p^2)
\label{selfenergy}
\earr
In perturbation theory, the bare (integer) ratio $q={4\pi m\over g^2}$
appearing in (\ref{condition}) acquires a multiplicative renormalization
\cite{pisarski}
\barr
q_{\rm ren}=4\pi \left [ {m\over g^2} \right]_{\rm ren} = {4\pi m\over
g^2}Z_m\tilde Z^2
\label{ren}
\earr
The multiplicative renormalization factor $Z_m$ is defined in terms of the odd
part of the gauge self-energy evaluated at zero momentum:
\barr
Z_m=1+\Pi_{\rm odd} (0)
\label{massren}
\earr
The renormalization factor $\tilde{Z}$ is similarly defined in terms of the
ghost self energy $\tilde{\Pi}$ \cite{pisarski}.

There are three types of contribution to the computation of $\Pi_{\rm odd}$, as
shown in Fig. 1. The first contribution, corresponding to the diagram in Fig.
1(a), arises from the external unbroken gluons coupling to internal unbroken
gluons, and so is exactly the same as that computed by Pisarski and Rao - so we
find
\barr
\Pi_{\rm odd}^{\rm Fig. 1(a)}(p^2)= 2 g^2 \int {d^3k\over (2\pi )^3} \left
[{p^2k^2-(p\cdot k)^2\over p^2}\right ] {5k^2 +5k\cdot p +4 p^2 +2m^2\over
k^2(k^2+m^2)(k+p)^2((k+p)^2+m^2)}
\label{graph1a}
\earr
where the overall factor of $2$ in front corresponds to the fact that this is
essentially an $SU(2)$ computation. At $p^2=0$ this reduces to
\barr
\Pi_{\rm odd}^{\rm Fig. 1(a)} (0) = 2 g^2 \left({2\over 3}\right) \int
{d^3k\over (2\pi )^3} {5k^2 +2m^2\over k^2(k^2+m^2)^2}\quad = \quad
 2 \;{g^2\over m} \;{7\over 12 \pi}
\label{graph1azero}
\earr

The second contribution to $\Pi_{\rm odd}$ comes from the diagram in Fig.1(b)
in which the external unbroken gluons couple to internal broken gluons, whose
propagator is now of the form (\ref{broken}), and so the Pisarski-Rao
computation must be modified. A lengthy, but straightforward computation, leads
to the expression
\barr
\Pi_{\rm odd}^{\rm Fig. 1(b)}(p^2)&=& {1\over 2}g^2 \int {d^3k\over (2\pi )^3}
{1\over Q} \left [{p^2k^2-(p\cdot k)^2\over p^2}\right ]\left\{
6m^2\right.\cr\cr
&&\left.\quad
-m^2\left(((p+k)^2+m_D^2)/(p+k)^2+(k^2+m_D^2)/k^2\right)\right.\cr\cr
&&\left.\quad -\left(6k^2+4p^2+6p\cdot k\right)
(k^2+m_D^2)((p+k)^2+m_D^2)/(k^2(p+k)^2)\right.\cr\cr
&&\left.\quad +(4p^2+4p\cdot k+8k^2)(k^2+m_D^2)/k^2\right.\cr\cr
&&\left.\quad +(8p^2+12p\cdot k+8k^2)((p+k)^2+m_D^2)/(p+k)^2\right\}
\label{graph1b}
\earr
where
\barr
{Q} \equiv (k^2+m_+^2)(k^2+m_-^2)((p+k)^2+m_+^2)((p+k)^2+m_-^2)
\earr
At $p^2=0$ this reduces to
\barr
\Pi_{\rm odd}^{\rm Fig. 1(b)}(0)&=& {1\over 3}g^2 \int {d^3k\over (2\pi )^3}
{6m^2k^2-2m^2(k^2+m_D^2)-6(k^2+m_D^2)^2+16k^2(k^2+m_D^2)\over (k^2+m_+^2)^2
(k^2+m_-^2)^2}\cr
&=&g^2 {1\over 2\pi (m_++m_-)}
\label{graph1bzero}
\earr
Expressions for this diagram Fig. 1(b) (at $p^2=0$) have been published
previously in \cite{khlebnikov2,khare1}. However these two previous expressions
disagree with one another, as well as with our result (\ref{graph1bzero}).
There is however, a check that when $m_D^2$ is set to zero inside the integral
appearing in the expression (\ref{graph1b}) (i.e. no symmetry breaking), this
should reduce (apart from the overall group theoretical factor of 2) to the
Pisarski-Rao expression in (\ref{graph1a}) for {\it all} $p^2$. This is true of
our expression (\ref{graph1b}), but it is not true if the coefficients are
modified to agree with the expressions in either \cite{khlebnikov2} or
\cite{khare1}. It is a remarkably unfortunate coincidence of numerology that if
one tries this $m_D^2\to 0$ reduction for the diagram evaluated at $p^2=0$,
then all three expressions reproduce the ${7\over 12 \pi}$ factor in the
Pisarski and Rao result (\ref{graph1azero}), {\it after} the $k$ integration
has been performed.

The third contribution to $\Pi_{\rm odd}$ comes from the diagram in Fig. 1(c)
in which the external unbroken gluons couple to an internal broken gluon and to
the internal unphysical scalar fields $\psi_1$, $\psi_2$, $\chi_1$ or $\chi_2$.
This diagram contributes
\barr
\Pi_{\rm odd}^{\rm Fig. 1(c)}(p^2)= g^2 m_D^2 \int {d^3k\over (2\pi )^3}
{p\cdot k\over p^2} {1\over (k^2+m_+^2)(k^2+m_-^2)(p+k)^2}
\label{graph1c}
\earr
which reduces at $p^2=0$ to
\barr
\Pi_{\rm odd}^{\rm Fig. 1(c)}(0)= g^2 {1\over 6\pi (m_++m_-)}
\label{graph1czero}
\earr

Combining these three results
(\ref{graph1azero},\ref{graph1bzero},\ref{graph1czero}) we find that the total
one-loop mass renormalization factor (\ref{massren}) is
\barr
Z_m&=&1+g^2\left [ 2\, {7\over 12\pi m} + {2\over 3\pi (m_++m_-)}\right ]
\label{zedm}
\earr

To compute $\tilde{\Pi}$ we need to compute the ghost self-energy, which has
two contributions, as shown in Figure 2. The contribution of Fig 2(a)
corresponds to an internal unbroken gluon, and so this is once again identical
to the Pisarski-Rao computation. This yields
\barr
\tilde{\Pi}^{\rm Fig. 2(a)}(p^2)= -2 g^2 \int {d^3k\over (2\pi )^3} \left
[{p^2k^2-(p\cdot k)^2\over p^2}\right ] {1\over k^2(p+k)^2(k^2+m^2)}
\label{graph2a}
\earr
At $p^2=0$ this reduces to
\barr
\tilde{\Pi}^{\rm Fig. 2(a)}(0)= -2 g^2 {1\over 6\pi m}
\label{graph2azero}
\earr
The second contribution comes from Fig. 2(b) in which the internal gluon
propagator is of the broken form (\ref{broken}), giving
\barr
\tilde{\Pi}^{\rm Fig. 2(b)}(p^2)= -g^2 \int {d^3k\over (2\pi )^3} \left
[{p^2k^2-(p\cdot k)^2\over p^2}\right ] {(k^2+m_D^2)\over
k^2(p+k)^2(k^2+m_+^2)(k^2+m_-^2)}
\label{graph2b}
\earr
which reduces at $p^2=0$ to
\barr
\tilde{\Pi}^{\rm Fig. 2(b)}(0)= -g^2 {1\over 3\pi (m_++m_-)}
\label{graph2bzero}
\earr
Thus the renormalization factor $\tilde{Z}$ is given by
\barr
\tilde{Z}=1- g^2 \left [ 2 {1\over 6\pi m}+{1\over 3\pi (m_++m_-)}\right]
\label{zedtilde}
\earr

We are now able to compute the renormalized ratio $q_{\rm ren}$ appearing in
(\ref{ren}):
\barr
q_{\rm ren}&=& {4\pi m\over g^2} \left\{ 1+g^2\left [ 2\, {7\over 12\pi m}
-4{1\over 6\pi m}+ {2\over 3\pi (m_++m_-)} -{2\over 3\pi (m_++m_-)}\right
]\right\}\cr
&=&{4\pi m\over g^2} \left\{ 1+2 {g^2\over 4\pi m} \right\}\cr
&=&q_{\rm bare} +2
\label{su3answer}
\earr
This shift by 2 corresponds precisely to the Pisarski-Rao shift for the
residual $SU(2)$ symmetry in the broken vacuum. For an $SU(N)$ theory
spontaneously broken to $SU(N-M)$, with $N-M\geq 2$, this computation goes
through almost entirely unchanged. For example, for $SU(N)$ broken to
$SU(N-1)$, the `additional diagrams' computed in
(\ref{graph1b},\ref{graph1c},\ref{graph2b}) are completely unchanged except for
the values of $m_\pm$. However, these cancel in the end and the $SU(N)\to
SU(N-1)$ generalization of the result (\ref{su3answer}) is
\barr
q_{\rm ren}&=& {4\pi m\over g^2} \left\{ 1+g^2\left [ (N-1)\, {7\over 12\pi m}
-2(N-1){1\over 6\pi m}+ {2\over 3\pi (m_++m_-)} -{2\over 3\pi (m_++m_-)}\right
]\right\}\cr\cr
&=&q_{\rm bare} + (N-1)
\label{sunanswer}
\earr

To conclude, we comment that this result fills a gap in our previous
understanding of the interplay between gauge invariance and the spontaneous
breaking of parity. The general picture in the abelian theories is well
understood - in the broken phase there is a radiative correction to the odd
part of the photon propagator \cite{khlebnikov1,spiridonov}, but this should be
understood not as a renormalization of the topological mass, but as due to the
appearance of gauge invariant terms in the effective action which reduce to a
Chern-Simons-like term when the scalar field takes its vacuum expectation value
\cite{khare2}. This then extends the Coleman-Hill theorem \cite{coleman},
concerning the absence of corrections to the topological mass in certain
abelian theories, to include the case of spontaneous symmetry breaking
\cite{khare2}. Presumably a similar mechanism operates in the completely broken
nonabelian theories \cite{khlebnikov2}, although this has not yet been
demonstrated explicitly. When a nonabelian symmetry is only partially broken,
leaving a residual nonabelian symmetry in the broken vacuum, this mechanism no
longer works - there is no obvious way to construct gauge invariant terms in
the effective action which could reduce to the nonabelian Chern-Simons term
when the scalar fields take their vacuum expectation value. However, the result
described here implies that no such extra terms are required. Our result should
also be relevant for the nonabelian generalizations of the abelian pure
Chern-Simons-matter theories considered in Ref. \cite{klee}.

\b\b

\noindent{\bf Acknowledgement:} We are grateful to Manu Paranjape for
discussions and correspondence. This work has been supported by the DOE grant
DE-FG02-92ER40716.00.

\begin{figure}
\caption{These diagrams represent the one-loop contributions to the odd part of
the (unbroken) gluon propagator $\Pi_{\rm odd}$. Figure 1(a) involves internal
unbroken gluon propagators, as in (\protect\ref{unbroken}), while Figure 1(b)
involves internal broken propagators, as in (\protect\ref{broken}). Figure 1(c)
involves an internal broken gluon propagator, and an internal unphysical scalar
propagator.}
\end{figure}

\begin{figure}
\caption{These diagrams represent the one-loop contributions to $\tilde\Pi$.
Figure 2(a) involves an internal unbroken gluon propagator, while Figure 2(b)
involves an internal broken propagator.}
\end{figure}

\end{document}